\documentstyle[preprint,aps,floats,epsf,epsfig]{revtex}
\begin{document}
%**************************************************************************
\thispagestyle{empty}
\title{ Complete  Leading Order Analysis in Chiral Perturbation 
Theory of the  Decays
$K_L\rightarrow \gamma\gamma$
and $K_L\rightarrow \ell^+ \ell^-\,\gamma$  }
%\vspace*{1.cm}\\
\author{J. L. Goity and Longzhe Zhang}
\address{
Department of Physics, Hampton University, Hampton, VA 23668, USA \\
and \\
  Jefferson Lab,  
12000 Jefferson Avenue, Newport News, VA 23606, USA.} 
 
\maketitle  
\begin{abstract}
The decays $K_L\rightarrow \gamma\gamma$
and $K_L\rightarrow \ell^+ \ell^-\,\gamma$ are studied at the leading order  
$p^6$ in Chiral Perturbation Theory. One-loop contributions stemming from the 
odd intrinsic parity  $\mid \Delta S\mid =1$ effective Lagrangian of order $p^4$
are included and shown to  be of possible relevance.  
They   affect the decay 
$K_L\rightarrow \gamma\gamma$ adding to the usual pole terms a piece
 free of counterterm uncertainties. 
In the case  of the   
  $K_L\rightarrow \ell^+ \ell^-\,\gamma$ decays the  
 dependence of the form factor on   the dilepton invariant 
mass requires a counterterm. The form factor may receive a sizeable
contribution from chiral logarithms. Including considerations from the  
$K_L\rightarrow \pi^+ \pi^-\,\gamma$ direct emission amplitude, 
we obtain 
two consistent scenarios.
  In one scenario the long distance contributions from the one-loop terms
   are   important,
while in the other they are   marginal. In both cases the counterterm is
shown to be significant.

\end{abstract}

\vspace{20mm}

\begin{flushright}
JLAB-THY-97-06
\end{flushright}

\newpage
 
 The radiative decays of the long lived neutral K meson,  
 $K_L$, have been the subject of numerous theoretical analyses,
 in   Chiral Perturbation Theory ($\chi$PT)$\cite{Dambrosio}$.
 Among these  decays, the decay $K_L \rightarrow \gamma\gamma$
   and the Dalitz decays $K_L \rightarrow  \ell^+\ell^- \gamma$ still 
   show open issues
   which we   address in this paper.  Both types of decays proceed
   at order $p^6$   in  $\chi$PT. The pole-type terms given by
   $\pi^0$ and the $\eta_8$ intermediate 
states$\cite{Gaillard,Donoghue,Goity}$ are individually
   of leading order $p^4$,  but their added contribution cancels at this order
due to the Gell-Mann-Okubo relation$\cite{Donoghue}$
 and is in effect  of order $p^6$. It is, therefore,
   important to perform a complete analysis of the decays at this order. We show that the
   one-loop terms of order $p^6$ stemming from the odd-intrinsic parity 
    terms of the $\mid \Delta S\mid =1$ non-leptonic effective Hamiltonian 
  of order $p^4$ and  not previously evaluated,  affect both types of    decays in a potentially 
 significant way.   We find, in particular, a contribution
to the amplitude
   of the two-gamma decay proportional to $M_K^2+M_{\pi}^2$ that can be
a substantial addition to the pole-type term.
In the case of the Dalitz decays,  the experimentally observed  non-trivial form factor
in terms of the dilepton invariant mass
is  usually described using   
  the VMD  model of Bergstr\"om, Mass\'o and Singer $\cite{Masso}$,
 which gives a satisfactory parameterization. 
The work of Ko$\cite{Ko}$ provides a thorough study of VMD in radiative decays.   
Recently,   other models  like
  the factorization model (FM) and the weak deformation model (WDM) 
   have been considered  as well$\cite{Portoles}$.  
Our calculation includes  long distance pieces given by  the chiral loops,
 and a short distance piece due to a counterterm
needed to renormalize the loops is determined by fitting to the data. 
   For our purpose, the experimental situation 
   is   good. The two-gamma mode is experimentally very precisely known:
    BR($K_L \rightarrow \gamma\gamma$)= $(5.92\pm 0.15)\times 10^{-4}$   
    $\cite{PDG}$. The branching ratios 
 of the Dalitz
 decays are on the other hand known within $10\%$:
  $BR(K_L \rightarrow   e^+  e^-\gamma)=(9.2\pm 0.5\pm 0.5)\times 10^{-6}$ 
(NA31$\cite{Kleinknecht}$ and BNL-E845$\cite{Ohl}$ collaborations),
   and $BR(K_L \rightarrow  \mu^+\mu^-\gamma)= (3.23\pm 0.23\pm 0.19) 
\times 10^{-7}$ (E799 collaboration$\cite{Enagonio}$). 
     The rate of the $\mu^+\mu^-$ mode is about $30\%$ larger than it would be
for the case of a point-like form factor. The form factor also shows up clearly in dilepton
invariant mass distribution of  both modes. Using this data and some input from
the $K_L\rightarrow\pi^+\pi^-\gamma$ decay$\cite{Neufeld}$, we are able
to  two phenomenologically acceptable scenarios.

   We work  in the standard framework of   $\chi$PT. The octet of pseudoscalar mesons
   is parameterized  by the SU(3) matrix 
     \begin{eqnarray}
    U&=& \exp \;( i \,\frac { \Pi  } { F_{ 0 } }  ),\nonumber\\
 \frac{1}{\sqrt{2}}\Pi & =& \left( 
\begin{array}{ccc}
 \sqrt{ \frac{1}{2} } \pi ^{0} + \sqrt{ \frac{1}{6} }\, \eta_8   & \pi^{+} & K^{+} \\
 \pi^{-}   & - \sqrt{ \frac{1}{2} } \pi^{0} + \sqrt{ \frac{1}{6} }\, \eta_8  & K^{0} \\
 K^{-}    & \bar{K} ^{0}  & -  \sqrt{ \frac{2}{3} }\, \eta_8 
\end{array}  \right) , 
\end{eqnarray}
where $F_0\sim F_{\pi}=93\; {\rm MeV}$ is the pion decay constant in the chiral limit.
For the processes we consider the only gauge couplings we need are those 
to the EM field. We thus denote:
 \begin{eqnarray}
 L_{\mu}&\equiv &i\; U^{\dagger}\,D_{\mu} \,U\nonumber\\
 D_{\mu}\,U&=&\partial_{\mu} U-i A_{\mu}\, [\hat{Q},\; U],
 \end{eqnarray}
where $\hat{Q}$ is the quark charge matrix.

   We need the strong interaction effective chiral Lagrangian  
     up to order $p^4$$\cite{GandL}$,
    including, of course, EM couplings and
   the Wess-Zumino (WZ) Lagrangian. The  non-leptonic 
   $\mid \Delta S=1\mid $  effective chiral Lagrangian  is   also required up 
   to order $p^4$$\cite{Kambor}$.
   It contains  an 8 and  a 27 of SU(3);  
  we advocate the
    $\Delta I=1/2$ rule to disregard  the marginal 
    effects of the 27 piece. 
      The octet piece of order $p^2$ is 
   \begin{equation}
  {\cal L}^{(2)}_{\mid \Delta S \mid = 1 } =
 \frac { F _ { 0 } ^ { 2 } } { 4 }\; C_ { 8 }\; 
 Tr ( \lambda_{6} \,L_{\mu}   \, L^{ \mu } ) ,
   \end{equation}
   where $C_8=3.12 \times 10^{-7}$ is given by 
  \begin{eqnarray}
C_{8} &=& 4 F_{0} ^{2} G_{8}  ,\nonumber\\
G_{8} &= & \sqrt { \frac { 1 } { 2 } } G_{F} \em s \/ _{1}
 \em c \/ _{1} \em c \/ _{3} g _{8} ,
\end{eqnarray}
 Here $s_i$, $c_i$ are the sine and the cosine of CKM angles in the 
 Kobayashi-Maskawa representation, and  $g _{8}\simeq 0.5$ is 
 an effective low energy coupling. 
 At order $p^4$ there  is a long list of terms (forty eight octet terms
   when external sources of different   kinds are included)
  $\cite{Kambor}$. The relevant terms to us are those  that affect the virtual
    transitions $K_L\rightarrow \pi^0,\; \eta_8$, including terms that  
    break SU(3) symmetry due to the quark masses, and
    terms that give the amplitudes $K_L\rightarrow \pi^+ \pi^-\gamma$ and 
    $K_L\rightarrow \pi^+ \pi^-\gamma\gamma$.
    The first  are contained in the terms bilinear in the meson fields of the  
${\cal O}(p^4)$ Lagrangian$\cite{Kambor}$, and the second are given by odd-intrinsic parity terms 
    of which only two combinations of the operators $O_i^8$ in$\cite{Kambor}$
     appear, 
    namely, 
    $W_{29}\equiv O_{28}^8+O_{29}^8$, and $ W_{31}\equiv O_{30}^8+O_{31}^8 $.
    The piece of the Lagrangian  ${\cal{L}}^{(4)}_{ \mid \Delta S = 1 \mid}$ 
    containing these two terms can be written as follows$\cite{Neufeld}$:

\begin{equation}
 {\cal{L}}^{(4)}_{ 29,\;31} = \frac{C_8}{64 \,\pi^2}\; \left\{  
a_2 \, {\rm Tr}(\Lambda\,[U^{\dagger} \tilde{F}^{\mu \nu}_R U,\;
  L_{\mu} L_{\nu}])+
  a_4 \, {\rm Tr} ( \Lambda\, L_{ \nu } ) \; {\rm Tr} ((\tilde{F} _{L} ^{ \mu \nu }
   - U^{ \dagger } \tilde{F} _{R} ^{ \mu \nu } U ) L_{ \nu })\right\}+{\rm h.c.}
\end{equation}
 where $\Lambda\equiv\frac{1}{2}(\lambda_6-i\lambda_7)$. 
 In our case, where the only external gauge field of relevance is the EM field, 
 we have ${F} _{R} ^{ \mu \nu }={F} _{L} ^{ \mu \nu }={F}^{ \mu \nu } \hat{Q}$,
 and $\tilde{F}^{ \mu \nu }\equiv \epsilon^{\mu\nu\rho\sigma} F_{\rho\sigma}$ is its dual. 
     Since the Lagrangians of order $p^2$
     are of even intrinsic parity,  they alone cannot generate 
      odd intrinsic parity  terms. Therefore,   the low energy couplings   $a_2$ and $a_4$
 are renormalization-scale independent. 
     
  In the absence of CP violation $K_L$ coincides with the CP-odd
   state $K_2=\frac{1}{\sqrt{2}}(K^0+\bar{K^0})$, and the pieces of ${\cal{L}}^{(4)}_{ 29,\;31}$ leading to the
  transitions  $ K_{L}\rightarrow \pi ^{+} \pi ^{-} \gamma $ and 
$ K_{L} \rightarrow \pi ^{+} \pi ^{-} \gamma \gamma$ are explicitly:
\begin{eqnarray}
{\cal {L}}_{ K_{L}\rightarrow \pi ^{+} \pi ^{-} \gamma }& = &
i \frac {e} { 8 \pi ^{2} F_{0}^{3} }\; 
C_{8}\; \epsilon ^{ \mu \nu \rho \sigma }\; 
[  a_{2}\, K_{L} \partial _{ \mu } \pi ^{+} 
\partial _{ \nu } \pi ^{-} \nonumber \\ &+ &
a_{4}\; \partial _{ \mu } K_{L} \;( \pi ^{-} 
\partial _{ \nu } \pi ^{+} - \pi ^{+} 
\partial _{ \nu } \pi ^{-} ) ] \;\partial _{ \rho } 
A _ { \sigma }  ~~, \nonumber \\
 {\cal {L}}_{ K_{L} \rightarrow\pi ^{+} \pi ^{-} \gamma \gamma   } 
&=& - \frac { e^{2} } { 8 \pi ^{2} F_{0}^{3} }\; C_{8} \;
\epsilon ^{ \mu \nu \rho \sigma }\; [ a_{2}\, K_{L} \,
\partial _{ \mu } A _{ \nu } \nonumber \\ 
&-& 2\, a_{4}\, \partial _{ \mu } K_{L} \,A _{ \nu } ]\; 
\partial _{ \rho } A _{ \sigma }\, \pi ^{+} \pi ^{-} +... 
\end{eqnarray}   
 In the second Lagrangian we do not display terms that do not contribute
 in our calculation.
     
  The SU(3) singlet   meson $\eta_{1}$,  being a massive 
    state due to the $U_A(1)$ anomaly,
    can be integrated out or, equivalently,     included explicitly
    in a meson pole dominance model$\cite{Donoghue,Goity}$. 
    The $\eta_{1}$ contribution to 
    the decays is naturally of order $p^6$. Since     $\eta_{1}$ and $\eta_{8}$
    mix with each other due to SU(3) breaking, we find it more convenient to include  
    $\eta_{1}$ explicitly.
    We need to consider both, 
    the terms that contribute to the virtual transition
    $K_L\rightarrow\eta_{1}$ 
    as well as   the WZ term, which leads to the standard amplitude for
     $\eta_{1}\rightarrow\gamma\gamma$.
    The   mixing angle between the $\eta$ mesons 
    and the decay constants as obtained from their two-gamma decays 
     and the ${\cal{O}}(p^4)$ result for
     $F_{\eta_8}$$\cite{GandL}$ are:
     \begin{eqnarray}
     \theta&=&-21\pm 5 \deg\nonumber\\
     F_{\eta_8}&=& 1.3 \, F_{\pi}\nonumber\\
     F_{\eta_1}&=&  1.1\,  F_{\pi}
    \end{eqnarray}
    The   virtual weak amplitude $K_L\rightarrow \eta_1 $ is not known, and 
    is characterized in the following by
    a parameter
     $\kappa$.

    In the limit of CP conservation, the $K_L\rightarrow\gamma\gamma^*$
    amplitude has the most general form:
    \begin{equation}
    A(K_L\rightarrow\gamma\gamma^*)= F(t)\;\epsilon^{\mu\nu\rho\sigma} \;
    \epsilon_{\mu} k_{\nu} \epsilon^*_{\rho} k^*_{\sigma}.
    \end{equation}
    Here $\gamma^*$ is a virtual photon, and $t=k^{* 2}$.
    The branching ratio for the two-gamma decay mode is  
    \begin{equation}
    B(K_L\rightarrow\gamma\gamma)=
    \frac{M_K^3 \mid F(0)\mid^2}{64 \pi \Gamma_{K_L}},
    \end{equation}
   and the dilepton distribution of the Dalitz decays is given by   
     \begin{equation}
         \frac{d\Gamma_{K_L\rightarrow \ell^+\ell^-\gamma}}{dt}=\
     \mid f(t)\mid^2 \Phi(t,\;m_\ell) 
     \;\Gamma_{K_L\rightarrow \gamma\gamma}, 
    \end{equation}
    where $f(t)\equiv F(t)/F(0)$,   $m_{\ell}$ is the  mass of the lepton, and
    \begin{equation}
    \Phi(t,\;m_\ell)\equiv  \frac{ \alpha_{\rm em}}{  t}
     \frac{2  }{3 \pi }\;(1-\frac{t}{M_K^2})^3 \;(1+\frac{2 m_{\ell}^2}{t})
    \sqrt{1-\frac{4 m_{\ell}^2}{t}} .
    \end{equation}

    There are two  different  sets of diagrams contributing to $F(t)$. 
    The first set are the pole terms shown in Figures 1 and 2 involving  the virtual transitions
    $K_L\rightarrow\pi^0,\;\eta_8,\;\eta_{1}$ followed by the two-photon
     decay modes driven by the WZ term.  The second set of diagrams,
     shown in Figure 2,
     requires the weak interaction transitions 
      $K_L\rightarrow\pi^+\pi^-\gamma$ and $ K_L\rightarrow
     \pi^+\pi^-\gamma \gamma$.
     
     As mentioned before,  the ${\cal O}(p^4)$ contributions due  to the $\pi^0$
     and $\eta_{8}$ poles cancel each other due to the Gell-Mann-Okubo (GMO) relation$\cite{Donoghue}$.
     At ${\cal O}(p^6)$      one-loop corrections   and
     counterterms must be considered.
     From this various effects result. One is due to the deviation from
     the GMO relation, which is taken into account by replacing the
     lowest order masses by the physical masses. Another one is due to  
     SU(3) breaking in virtual amplitudes $ K _{L} \rightarrow \eta _{8}  $ 
 with respect to $  K _{L} \rightarrow  \pi^0  $, which we take into account by means of
 the parameter $\delta$ below.    
     A final effect stems from the one-loop corrections to the WZ
       term. As   shown in$\cite{Bijnens}$, there is a $t$ independent piece
       that is taken into account by simply replacing the decay constants
       in the chiral limit by the physical decay constants, 
       while the $t$ dependent piece
       requires renormalization. Fortunately,  the latter
       only appears in contributions to the $K_L\rightarrow \ell^+\ell^-\gamma$ 
       decay   of order $p^8$, and can therefore be altogether disregarded.
       The reason is that the $t$-dependent  corrections to the WZ 
       term have the same relative weight for  the
        $\pi^0$ and $\eta_8$ as in the WZ term itself$\cite{Bijnens}$, and therefore 
        the mechanism of cancellation  by the GMO relation    eliminates  the order $p^6$ terms.
 Since the $\eta_1$ contribution is 
        already of order $p^6$ at tree level, the one loop corrections
        are of order $p^8$ and, therefore, disregarded. From this, 
        the  pole piece of $F(t)$  can be expressed as follows$\cite{Donoghue,Goity}$:
        \begin{eqnarray}
          F_1  &=& -\frac{\alpha_{\rm em}}{2 \pi F_{\pi}} \; C_8  \; \tilde{F_1}\nonumber\\  
    \tilde{F_1} &\equiv&  r_{ \pi }   + r_{ \eta }  \Theta_{1} +
     r_{\eta^{ \prime } }  \Theta_{2}  ,
        \end{eqnarray}
     where         
      \begin{eqnarray}
    r_{  P } &\equiv & 1/\left(1 - \frac { m _{P }^{2} } { M_{K_L} ^{2} }\right),
    \nonumber \\
\Theta _{1} &= &\frac {1} {3}\; \left((1 + \delta )\, \cos \theta +
 4   \, \kappa \,\sin \theta \right) \;
 \left( \frac { F_{ \pi } } { F_{ \eta _{8} } } \,\cos \theta - 
 2 \sqrt {2} \frac { F_{ \pi } } { F_{ \eta _{1} } } \,\sin \theta \right), \nonumber \\
\Theta _{2}& =& \frac {1} {3} \;\left( ( 1 + \delta )\,\sin \theta - 
4 \,  \kappa\,\cos \theta \right)\;
\left( \frac { F_{ \pi } } { F_{ \eta _{8} } }\,\sin \theta + 
2 \sqrt {2}\; \frac { F_{ \pi } } { F_{ \eta _{1} } }\,\cos \theta \right).
\end{eqnarray}
We emphasize that $F_1$ 
 is very sensitive to the  parameteres $\kappa$ and $\delta$, and therefore, cannot be reliably predicted.
      
 The one-loop diagrams of Figure 2 containing the vertices of order $p^4$ in
  (6) give the  following 
contribution to $F(t)$:  
\begin{equation}
  F_{2} (t) =\frac { \alpha_{\rm em} C_{8} } { 192 \pi ^{3} F_{0} ^{3} } \,
   \left\{(a_{2} + 2 a_{4} )\; 
 \left[ 16 (M_{\pi} ^{2} + M_{K} ^{2} ) - D (t,\mu) \right]+CT(t,\mu)\right\}~~~,
\end{equation}
where CT denotes a counterterm, and the function $D (t,\mu)$ is given in dimensional regularization  by
    \begin{eqnarray}
    D(t,\mu)=t\; [\frac{10}{3}+2\lambda- 
     ( \log{\frac { M_{K} ^{2} } { \mu ^{2} }} +
      \log{\frac { M_{\pi} ^{2} } { \mu ^{2} } })]     \nonumber \\ 
      + 4 ( F( M_{\pi} ^{2}, \em t ) + F( M_{K} ^{2}, \em t) )  ~~,
 \end{eqnarray}
where
\begin{eqnarray}
F( m ^{2}, ~t) &\equiv & \left(( 1 - \frac {y} {4} )\; 
\sqrt { \frac {y - 4} {y} } \log{\frac { \sqrt {y} 
+ \sqrt {y  - 4 } } { - \sqrt {y } + \sqrt {y - 4 } }}
 - 2 \right) m^{2} ,\nonumber \\ 
  y  &\equiv & \frac{t}{ m^{2}} ,
   ~~\lambda \equiv \frac {1} {  \varepsilon } + 1 + \log 4 \pi - \gamma _{E} .
\end{eqnarray}
U-spin symmetry of  ${\cal L}_{\mid \Delta S=1\mid }$  
implies that the 
$\pi^+$ and the $K^+$  loops  have the same sign.

   Only the $t$-dependent piece 
     needs renormalization   provided by  the counterterm  $CT(\mu)$.   
The t-independent UV divergencies cancel when the two diagrams in Figure 2 are added.
The t-dependent piece needs  
       renormalization  that is, as usual, accomplished by replacing in (14) $CT(t, \mu)$ plus the term
     proportional to $\lambda$ by    $C(\mu)\,t$. As   shown later,
      this counterterm can be estimated by analyzing  the data. Therefore,      
$F_2(0)$ is unaffected by counterterm uncertainties, and is
      non-vanishing   and
       proportional to $M_K^2+M_{\pi}^2$. This term is thus a new
     addition to the $K_L \rightarrow \gamma\gamma$ amplitude. 
Notice that one can write a finite  counterterm for the $t=0$ piece. Such a counterterm
is included in $F_1$, and in the pole model it is   given by the $\eta_1$ intermediate state as 
already discussed. The situation here
resembles that of the strong interaction case at order $p^4$, where the term proportional to
the low energy constant $L_7$  only provides a finite renormalization, and is in fact dominated by the
$\eta_1$ pole$\cite{GandL}$.  
     The $t$ independent pieces of $F(t)$ vanish  in the chiral limit
     in accordance with the Veltman-Sutherland theorem.

  The unknown parameters affecting the two-photon and Dalitz modes  are $\tilde{F}_1$ and
  $(a_2+2 a_4)$, while the counterterm $C(\mu)\,t$
    only affects  the Dalitz mode. 
     When we  neglect  the 
  t-dependence in $F_2$,    (10) gives  
  \begin{equation}
  \Gamma_{K_L\rightarrow \ell^+\ell^-\gamma}^{(0)}=  \Gamma_{K_L\rightarrow \gamma\gamma}
  \int_{4 m_{\ell}^2}^{M_K^2} \Phi(t,\;m_{\ell})\;dt 
  \end{equation}
  and the ratios with the experimental values  are   $\cite{PDG}$:
     \begin{eqnarray}
 R_{e^+e^-\gamma}\equiv    \frac{\Gamma_{K_L\rightarrow e^+e^-\gamma}^{Exp}}
{\Gamma_{K_L\rightarrow e^+e^-\gamma}^{(0)}}&=&0.97\pm 0.05\nonumber\\\nonumber\\
  R_{\mu^+\mu^-\gamma}\equiv      \frac{\Gamma_{K_L\rightarrow \mu^+\mu^-\gamma}^{Exp}}
{\Gamma_{K_L\rightarrow \mu^+\mu^-\gamma}^{(0)}}&=&1.36\pm 0.14
      \end{eqnarray}
  Thus,  (17) is an excellent approximation  in the $e^+e^-$  mode, where most of the 
  rate is given by the  low-t domain  where the relative   t-dependence is very small,
     while in the  
   $\mu^+\mu^-$ case   
      $t_{\rm min}=4 m_{\mu}^2$  is large enough for the $t$-dependence to be
   noticeable. Thus,   on the basis of the total rate of the 
 $\mu^+\mu^-$ Dalitz mode it is clear that
   there is a significant $t$-dependence  in $F_2$, 
as was first experimentally noticed by the E799 Collaboration
  $\cite{Enagonio}$. Denoting for the sake of convenience,
   \begin{eqnarray}
 C_1& \equiv & (a_2+2 a_4),   \nonumber\\
   C_2&\equiv& \frac{1}{4}\;(a_2+2 a_4)\;
   \left(\frac{10}{3} -(\log\frac{M_K^2}{\mu^2}+
   \log\frac{M_{\pi}^2}{\mu^2})\right)-\frac{C(\mu)}{4}~~,
 \end{eqnarray} 
  from  the ratio $R_{\mu^+\mu^-\gamma}$ we find that they are 
approximately 
linearly related as 
   \begin{equation}
      C_2\simeq - {\rm sign} (F(0))\,(3.8\pm 1.3)+ 1.1\, C_1. 
   \end{equation} 
With this relation we   also find an excellent fit to both  dilepton invariant mass 
distributions$\cite{Kleinknecht,Ohl,Enagonio}$. As expected,   the
Dalitz decays  alone cannot  pin-down    $C_1$ and the counterterm.
It is  necessary to invoke further observables for this to be possible. 
 The value   of $C_1$ can also be estimated 
by means of two other processes, namely,    
$K_L \rightarrow \gamma\gamma$ and  $K_L \rightarrow \pi^+ \pi^- \gamma$.
The amplitude of the first process is given in terms of 
$F(0)=\frac{\alpha_{\rm em} C_8}{2 \pi F_{\pi}}\;L_1$ with  
\begin{equation}
L_1\equiv -\tilde{F_1}+C_1 \;\frac{M_K^2+M_{\pi}^2}{6 \pi^2\,F_{\pi}^2}. 
\end{equation} 
From the $K_L \rightarrow \gamma\gamma$ width (9) this combination has the 
value $\pm 0.89$.
On the other hand, the second process permits a model dependent 
 estimate of a different combination$\cite{Neufeld}$, namely,
\begin{equation}
L_2 \equiv C_1 - \tilde{F_1}. 
\end{equation} 
While in  (21) both terms   are of the same chiral order, in  (22) 
the first term 
is of lower   order than the second. 
Indeed, $C_1$ gives the strength of the direct 
emission
$M1$ amplitude of order $p^4$, while the second term appears in the 
correction of order $p^6$ 
 to that amplitude$\cite{Neufeld}$.
 There is, in addition, the theoretically well known internal bremsstrahlung amplitude of order $p^2$
that is CP violating  
and  cannot interfere with the direct emission one
because it is of electric type.  Thus, the  direct emission decay rate can be cleanly
identified, and
 the combination in (22) is  then estimated to
be in the interval 0.3 to 0.9 $\cite{Neufeld}$. The positive sign of the combination
is favored by a factorization model$\cite{Neufeld}$, but the possibility 
of a negative sign is not altogether ruled out. For a similar negative interval the 
conclusions we draw here remain unchanged. 
Taking the estimated range at face value,   we  
 obtain $C_1$ and $\tilde{F_1}$ for the two possible signs of $L_1$ as shown in the table.
A first consequence is that the amplitude for the 
two-gamma decay has a potentially important  addition  beyond the pole terms, depending
on the scenario, defined by the sign of $ L_1$,  one considers. 
This is shown in the table by the ratio $F_2(0)/F_1$, which can be as large as  $0.6$.

The slope of the Dalitz decay form factor defined by
\begin{equation}
b\equiv \frac{1}{2} \; \frac{d}{dx}  f^2(t)   \mid_{x\rightarrow 0}=
\frac{  M_K^2}{24\,\pi^2\,F_{\pi}^2\, L_1} \;(\frac{4}{3} C_1-C_2),~~x\equiv \frac{t}{M_K^2} 
\end{equation}
is consistent with the experimental value obtained from the $e^+e^-$ mode, 
$b_{\rm Exp}=0.6\pm 0.25$$\cite{Kleinknecht,Ohl}$. In fact, in the acceptable range of $C_1$ (20) gives
$b=0.45\pm 0.2$. We notice that the chiral logarithm terms as well as the counterterm  give positive contributions to
the slope. The fraction of the slope due to the counterterm 
is shown in the last entry of the table.  We see that in one case
it is dominating, giving  support to the VMD model$\cite{Masso}$, 
while in the other case there are important 
chiral logarithm terms as well. 
 
 Since the $\mu^+\mu^-$ mode shows 
more prominently     the t-dependence, the results are mostly determined by that mode.
In turn, for the $e^+ e^-$ mode we predict   $R_{e^+e^-\gamma}=1.025\pm 0.010$.
This ratio is consistent within the $5\%$ error of the
 experimental result (18).

In conclusion, our leading order analysis shows two acceptable scenarios. In one of them
there are sizeable long distance contributions from one-loop diagrams to both 
 types of radiative decays considered, while  in the other 
  such contributions are small,  implying that VMD models are an excellent picture.
 The question of   which of the two scenarios is actually realized  is clearly important.

\section{Acknowledgements}
 
We thank  K. Kleinknecht and J. Scheidt  for some useful correspondence about data.
  JLG is partially supported  by NSF grant HRD-96-33750.

\def\prd#1#2#3{{\it Phys. ~Rev. ~}{\bf D#1} (19#2) #3}
\def\prc#1#2#3{{\it Phys. ~Rev. ~}{\bf C#1} (19#2) #3}
\def\plb#1#2#3{{\it Phys. ~Lett. ~}{\bf B#1} (19#2) #3}
\def\npb#1#2#3{{\it Nucl. ~Phys. ~}{\bf B#1} (19#2) #3}
\def\npa#1#2#3{{\it Nucl. ~Phys. ~}{\bf A#1} (19#2) #3}
\def\prl#1#2#3{{\it Phys. ~Rev. ~Lett. ~}{\bf #1} (19#2) #3}

\bibliographystyle{unsrt}

\newpage\thispagestyle{empty}

\begin{center}
{\bf FIGURES}
 \end{center}

 \begin{center}
\vspace*{1in}
\vbox{\centerline{\epsfig{file=dKLfig1.eps ,height=4.0cm}}}
\vspace*{7 mm}
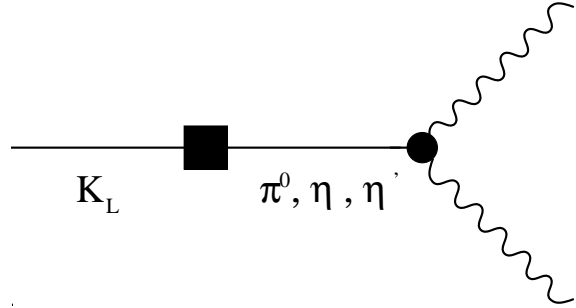
\parbox{4.5in}{{\bf Figure 1:}  {Pole diagrams.
The square represents the insertion of the $\mid \Delta S=1\mid $ effective non-leptonic weak interaction and the dot
is the WZ term.    }} 
\vspace*{7 mm}
\end{center}

%\begin{center}
%\vspace*{0.2in}
%\vbox{\centerline{\epsfig{file=dKLfig2.eps ,height=10.0cm}}}
%\vspace*{9 mm}
%\parbox{4.5in}{{\bf Figure 2:}  {  Diagrams of order $p^6$. }} 
%\vspace*{7 mm}
%\end{center}

\begin{center}
\vspace*{0.5in}
\vbox{\centerline{\epsfig{file=dKLfig3.eps ,height=6.0cm}}}
\vspace*{9 mm}
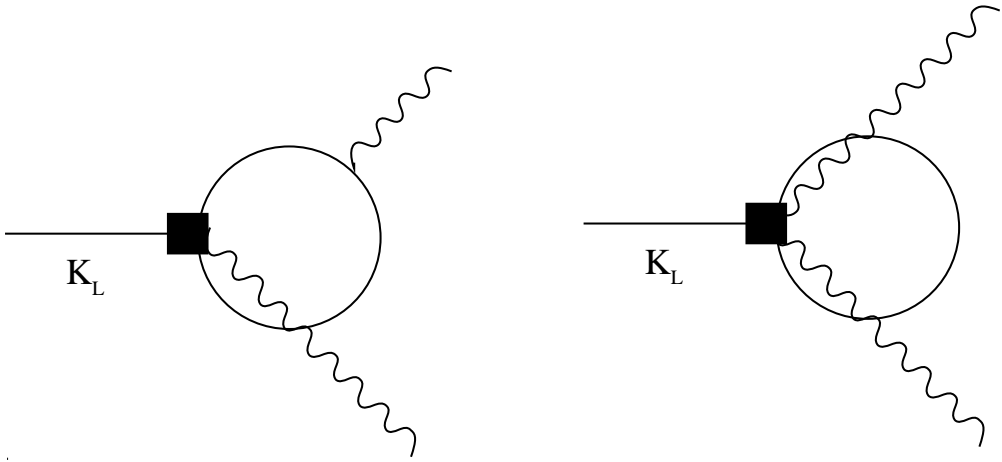
\parbox{4.5in}{{\bf Figure 2:}  {  Order $p^6$ one-loop diagrams. The square represents
the insertion of ${\cal{L}}^{(4)}_{29,~31}$. The mesons in the loop are  $\pi^+$ and $K^+$. }} 
\vspace*{7 mm}
\end{center}

 \newpage\thispagestyle{empty}
\begin{center}
{\bf TABLE}
  
\vspace*{1in}

\begin{tabular}{|c|c|c|c|c|c|c|} \cline{1-7}
$L_1$& $C_1$ & $\tilde{F}_1$ & $C_2$   
& $F_2(0)/F_1$ & $b$ & $ b_{\rm CT}/b$ \\
 & & & & & & \\ \hline
0.89 & & & & & &\\
$~0.3\leq L_2 \leq 0.9~$ & $~-0.6\pm 0.6~$ 
& $-1.2\pm 0.3~$ &  $-4.5\pm 1.3~ $   
& $~0.25\pm 0.25~$& $~0.5\pm 0.2 $ ~&$~0.9\pm 0.1~$ \\
& & & & & & \\ \hline
$-0.89 $ & & & & & &\\
$0.3\leq L_2 \leq 0.9$ & $~3.1\pm 0.6~$ & $~2.5\pm 0.3~$ &  $7.2\pm 1.3$  
  & $0.65^\pm 0.15$ & $ ~ 0.4\pm 0.2 ~  $ & $~0.4\pm 0.1~$ \\
& & & & & & \\ \hline
\end{tabular}
 \end{center}
\end{document}